\documentstyle[12pt,aasms4]{article}

\begin{document}

\title{FUV Spectroscopy of the Supersoft X-ray Binary RX~J0513.9$-$6951
\footnote{Based on observations made with the NASA-CNES-CSA Far
Ultraviolet Spectroscopic Explorer.  FUSE is operated for NASA by the
Johns Hopkins University under NASA contract NAS5-3298} } 

\author{J.B. Hutchings, K. Winter}
\affil{Herzberg Institute of Astrophysics, NRC of Canada,\\ Victoria, B.C.
V8X 4M6, Canada; john.hutchings@nrc.ca} 

\author{A.P. Cowley, P.C. Schmidtke}
\affil{Department of Physics \& Astronomy, Arizona State University,
Tempe, AZ, 85287-1504; anne.cowley@asu.edu; paul.schmidtke@asu.edu } 

\and
 
\author{D. Crampton}
\affil{Herzberg Institute of Astrophysics, NRC of Canada,\\ Victoria, B.C.
V8X 4M6, Canada} 

\begin{abstract}

We have obtained spectroscopy with the Far Ultraviolet Spectroscopic
Explorer (FUSE) of the supersoft X-ray binary RX~J0513.9$-$6951 over a
complete binary orbital cycle.  The spectra show a hot continuum with
extremely broad O VI emission and weak Lyman absorptions.  He II emission
is weak and narrow, while N III and C III emissions are undetected,
although lines from these ions are prominent at optical wavelengths.  The
broad O VI emission and Lyman absorption show radial velocity curves that
are approximately antiphased and have semiamplitudes of $\sim117\pm40$ and
$54\pm10$ km s$^{-1}$, respectively.  Narrow emissions from He II and O VI
show small velocity variations with phasing different from the broad O VI,
but consistent with the optical line peaks.  We also measure considerable
changes in the FUV continuum and O VI emission line flux.  We discuss the
possible causes of the measured variations and a tentative binary
interpretation. 

\end{abstract} 


\section{Introduction}

In this paper we present FUSE observations of the supersoft X-ray binary
RX~J0513.9$-$6951 (hereafter called X0513$-$69).  This $ROSAT$ source
was identified with a $\sim$16th mag peculiar emission-line star in the
Large Magellanic Cloud (Pakull et al. 1993, Cowley et al. 1993).  With an
absolute magnitude of M$_V\sim-2.0$, X0513$-$69 is the brightest of the
supersoft X-ray binaries (e.g. Cowley et al. 1998).  It is known to show
high and low optical states differing by $\sim$1 mag (e.g. Alcock et al.
1996, Cowley et al. 2002).  The optical spectrum is characterized by
strong, broad emission lines of He II and weaker lines of O VI, N V, C IV,
and C III.  Weak emissions flanking the strongest lines at $\pm\sim$4000
km s$^{-1}$ have been interpreted as arising in bi-polar jets (e.g.
Crampton et al. 1996). 

Previous spectroscopic work has shown that the emission line peaks have
very small velocity amplitude (e.g. Crampton et al. 1996, Southwell et al.
1996).  If these are interpreted as motion of the compact star, the source
must be seen at a low orbital inclination and the unseen secondary star
must have a low mass.  The small amplitude of the optical light curve
(Alcock et al. 1996) also suggests the system is viewed from a low
inclination angle.  However, using a newly refined orbital period and
ephemeris, Cowley et al.\ (2002) have shown that the phasing of the
velocities derived from optical spectra differs between the high and low
optical states, so that the previously inferred mass function and
its interpretation are in doubt. 

In order to access other highly ionized lines (particularly the O VI
resonance doublet) and investigate the far-ultraviolet continuum,
observations were obtained with the Far Ultraviolet Spectroscopic Explorer
(FUSE).  The FUSE data presented in this paper add new information about 
X0513$-$69. 

\section{Data and Measurements}

The FUSE observations were taken within a 24 hour period on 2001 October
30.  The windows during which the system can be observed, between earth
occultations and SAA passes, allowed 16 exposures ranging from 367 to 3310
seconds, spread fairly evenly over $\sim$1.2 binary cycles.  The data were
processed using CALFUSE version 2.0.5.  The source is bright enough so
that there were no difficulties with background subtraction or extraction
windows.  Analysis of the FUSE FES images, taken during the acquisition
and observation sequence, shows that X0513$-$69 was in its high optical
state at 16.7$\pm0.2$ mag during the far ultraviolet observations. 

In general, the LiF1a channel values are the most reliable for wavelength
and flux, as this is the guide channel.  LiF1b is usually well-tracked
too, as it suffers little thermal cycling.  However, as the observations
were continuous, the stability of all the telescope alignments is likely
to be good, and the data rates for all channels showed no signs of losing
the target star. 

Figure 1 shows the mean FUSE spectrum of X0513$-$69, with various features
marked.  We note the lack of N III or C III lines, although they are
present in optical spectra.  The principal region of interest is the O VI
doublet ($\lambda$1032,1038), seen as a broad emission blend, but
contaminated by some sharp airglow lines.  No broad emissions were seen at
other less ionized lines, such as He II, C III, or N III.  There are also
weak sharp emissions at the positions of the O VI lines, unfortunately
blended with airglow and interstellar absorption features.  However, the
relative strengths of these and other unblended airglow lines indicate
that the sharp O VI peaks are present, superimposed on the broad emission.
We note that the weak O VI lines in the visible region (Crampton et al
1996) are seen only as
sharp peaks, with no broad component detected.  In the FUSE spectra He II
is only detected as a sharp emission peak, weak and blended with airglow
emissions.  By contrast, both a strong narrow component and broad emission
wings are seen in the He II line at 4686\AA.  In HST spectra, G\"ansicke et
al.\ (1998) found a similarly broad structure in the N V resonance doublet 
(1239,1243 \AA), also with narrow peaks (see their Figs.\ 1 \& 4).  The
continuum is present throughout the entire FUSE range, and there are
narrow Lyman absorptions, seen most clearly in lines L$\delta$ through
L$\zeta$. 

Table 1 gives the observational details and some of the measurements made.
In all cases, the new ephemeris by Cowley et al.\ (2002) is used.  The
eight different FUSE spectral channels were measured separately.  Those
with the best signal level are reported here.  Measurements of the FUSE
spectra were made by fitting single profiles to the O VI emission blend,
and also by cross-correlation against a smoothed mean spectrum template,
which had been edited to remove airglow emission and narrow interstellar
absorptions.  Airglow emissions were identified using the atlas in Feldman
et al.\ (2001) and also by comparing spectra with varying fractions of
nighttime in the exposure.  The Lyman absorption lines were measured by
simple profile fitting.  The continuum variations were measured by
averaging the total fluxes outside of the emission features and with
airglow emissions removed.  Figure 2 shows plots of the measured
quantities, and Figure 3 shows the O VI profile in detail and a montage 
of changes with orbital phase. We discuss the broad O VI in detail in
the following section. 

Since there was a large range in individual exposure times among the
spectra listed in Table 1, we generated a set of 9 spectra combined to
equalize their exposures and phase coverage, and hence enhance the S/N for
measurement of the weaker features.  These were also used as a check for
our major results.  In all cases these phase-averaged spectra gave similar
results, so we do not quote these measures in detail.  We note, however,
that there are differences in some measures between the overlapping phases
from the two orbital cycles covered.  Further observations would be
required to fully separate the phase-dependent and irregular changes. 

All measurements were run through a sine-wave fitting program to determine
relative binary phases of their variations, using the Cowley et al.\
(2002) ephemeris.  The results and their formal errors are shown in Table
2.  Inspection of Figure 2 shows that some variations are not very
sinusoidal, especially the velocity curves of the broad O VI emission and
Lyman absorption, but the formal errors of the fits are a useful measure.
Using more complex functions, such as elliptical orbit fits, does not
alter the amplitudes and phasing significantly.  Note that the phase
errors are all quite small, with the exception of He II emission
velocities. 

\section{O VI Broad Emission} 

The O VI emission is extraordinarily broad and cannot be separated into
the two O VI lines at 1032 and 1038\AA.  Moreover, the whole emission
feature is redshifted by some 800 km s$^{-1}$ or more ($\sim$3\AA) from
the mean wavelength expected for the blend at the velocity of the LMC. 
The profile is complicated by the presence of strong C II 1036\AA\ and
other weaker interstellar absorptions, and by airglow emissions, including
the strong Ly$\beta$ line at 1026\AA.  Figure 3 shows the mean O VI
profile and its variation with phase. 

It appears that the emission actually extends to $\sim\pm4000$ km
s$^{-1}$, but there is a P Cygni absorption that erodes the short
wavelength side.  The amount of O VI absorption is evident from the
difference between the profile and its reflection about zero velocity
which is shown in Figure 4.  Zero velocity is taken as the mean
wavelength of the two O VI lines, shifted by the systemic velocity
($\sim$+280 km s$^{-1}$).  This assumes equal flux from each line (i.e.
high optical depth).  This means that the velocity width for each
individual line is $\sim$3000 km s$^{-1}$. 

The high velocity optical jets seen in H and He II lines (Crampton et al.,
Southwell et al.) do not appear to have a counterpart in the far
ultraviolet O VI lines.  In the optical region, the jet lines stand well
above the broad wings in most observations, while Figure 4 shows a smooth
profile with no features at $\pm4000$ km s$^{-1}$.  However, we note that
the optical jet lines are variable from epoch to epoch.  Cowley et al.\
(1998) found the jet lines very weak or absent in spectra taken in 1996
during a bright optical state, whereas they had been detected easily in
1993 and 1994 spectra (see their Fig.\ 3).  The broad O VI emission
profiles do extend out to these high velocites, as do the He II 4686\AA\
wings, but the 4686\AA\ wings could not be measured accurately for radial
velocity changes because of blending with the nearby N III/C III blend. 
However, Crampton et al.\ did report extended shortward absorption in the
broad wings of H$\beta$, indicating a similar P Cygni wind profile as seen
in O VI (see Figure 4). 
 
There are at least three possible causes of the measured broad emission
radial velocity variations.  These include: (a) changing strength of the P
Cygni absorption, altering the mean position of the unabsorbed emission,
(b) variation of the relative strengths of the two O VI doublet lines,
corresponding to changes in the optical depth of the emission region, (c)
velocity variations which reveal orbital motion of the emission line
region. 

Both possibilities (a) and (b) above would involve changes in the overall
profile of the emission feature.  Crampton et al.\ report that in their
1994 data ``the H$\beta$ wind absorption appears to be strongest at phase
0.75'' (which is phase 0.72 using the revised ephemeris of Cowley et al.
2002), but from limited data they also note that the ``changes are of low
significance and may not be phase related''.  This phasing corresponds
roughly to the maximum velocity of the broad O VI emission, suggesting the
velocity variation might be related to changing P Cygni absorption strength.
However, in the FUSE broad O VI profiles we do not see such a change, but
rather only changes in the overall emission flux, (plus associated changes
in the interstellar absorption depths).  Profile changes that give rise to
the measured full velocity amplitude of $\sim$240 km s$^{-1}$ would show
up as a profile difference, significantly different on the opposite sides
of the central wavelength.  This is not seen at any phase, and
particularly not between the phases of maximum velocity differences, where
we see the expected `sine-wave' difference spectrum from a moving but
unchanging profile.  We thus consider that the broad profile arises in a
region that appears similar from all viewing angles. 

The phasing of the measured velocity curve is close to that expected for
the compact star's orbital motion, assuming the optical light minimum
occurs near its superior conjunction (Cowley et al. 2002).  Thus, we
consider the implications of possibility (c).  The small velocity
amplitude of the peak emission velocities were discussed in this context
by Crampton et al.\ and Southwell et al.  Their interpretation implies
either a low orbital inclination and a low-mass donor star
($\sim$0.3M$_{\odot}$).  However, as shown by Cowley et al.\ (2002), the
new ephemeris makes this model unlikely, since maximum velocity does not
occur at the expected phase.  Therefore, we explore the possibility that
the $\sim$120 km s$^{-1}$ semiamplitude of the O VI broad emission is an
upper limit to the compact star's motion.  If there are small profile
changes that contribute to the velocity amplitude, they will reduce the
contribution from the orbital motion.  Thus, this measured velocity gives
a maximum value to the mass function for the system.  We discuss the
implications in more detail below. 

The very broad O VI emission most likely arises in the innermost parts of
the accretion disk, as this is where we expect the most highly ionized
material to be found, if radiatively excited.  We argue below that the
mass of the compact star probably lies in the range 0.7 to 1.1M$_{\odot}$,
while the orbital inclination must be $\sim35^{\circ}$, given the line
widths.  The orbital velocity at the surface of such a white dwarf is in
the range 3100 to 5500 km s$^{-1}$, which would be projected to 1900 to
3400 km s$^{-1}$ for an equatorial disk that reaches the stellar surface. 
The emission profile from such a disk at this inclination angle would not
be strongly double-peaked and the blend of two O VI line profiles
(separated by some 1700 km s$^{-1}$) together with interstellar
absorptions, would further obscure any profile structures. 

An alternative site for high velocities in the O VI profile is the
bi-polar jets.  As noted, the jets are seen in much lower ionization lines
as two separate and relatively narrow peaks, implying a high degree of
collimation.  The O VI could conceivably arise in the inner accelerating
part of the jets.  However, since the jets are highly collimated (the jet
emissions indicate an opening angle less than 10$^{\circ}$), this would
give rise to very double-peaked structure at any point in the jet, quite
unlike the profiles seen in the FUSE data. 

Given the probable origin of the broad emission in the inner disk, and its
unchanging profile with phase, we note that the P Cygni absorption must
indicate a considerable (and azimuthally symmetrical) disk wind that is
seen around the viewing cone and which must be separate from the jets.
Absorption within a jet would only be seen in a narrow viewing cone of
about half the 10$^{\circ}$ jet opening angle. 

\section{FUV Light Curve and Other Line Features}

The continuum is detected throughout the FUSE spectral range.  Continuum
variations were measured by averaging the total fluxes in regions outside
of the emission features where the airglow emissions had been removed. 
The measured flux changes are shown in Figure 2 and in Tables 1 and 2.
Table 1 shows values of the interpolated continuum at the O VI doublet,
while the values in Table 2 and Figure 2 are continuum levels fitted
uniformly to the wavelength regions indicated.  The agreement between the
independent FUSE channels is very good.  Note the rising flux level to
shorter wavelengths, indicating very high temperatures.  Expressed in
magnitudes, the orbital variation is $\sim$0.5 -- 0.6 mag, but examination
of overlapping phases shows this amplitude is partially due to
cycle-to-cycle variations.  The FUV variation is much larger than that
seen in the optical, where the mean range is only $\sim$0.04 mag, but the
cycle-to-cycle variations are up to $\pm0.2$ mag (see Cowley et al. 2002).

The FUV continuum and its changes must come from a region with
temperatures of 30000 K or more, to be so much larger in the FUV.  We also
note that the O VI broad emission flux varies with similar amplitude and
phasing, so it is likely to be associated with the continuum changes.
Since disk light dominates the spectrum, we are seeing azimuthally varying
disk brightness.  One possibility is a thickening of the outer disk
downstream of the gas stream from the donor star.  Illuminated by the hot
inner disk, this would be brightest when viewed from the opposite side. 
We discuss below a model where the phasing of this is consistent with the
radial velocity changes. 

Cowley et al.\ (2002) discuss the origin of the optical light variations,
showing the phasing of the minima has persisted for at least 8 years.
Thus, it is likely to be caused by a geometric occultation of a bright
region fixed within the system, rather than disk structure whose position
and shape may change.  This region may be the inner point in the Roche
lobe, where heating by the WD and X-rays is greatest.  Thus, they suggest
that phase 0 is at or close to superior conjunction of the compact star.
We adopt this geometry in our binary discussion below. 

It was not possible to make clean measures of the O VI peak emissions
because of blending with the airglow and interstellar lines.  The 1038\AA\
line velocity shows a correlation with the fraction of the exposure that
was in daylight (i.e. strength of airglow).  If we fit a straight line
through this, we find a velocity of +295 km s$^{-1}$ for an airglow-free
spectrum, which is close to the systemic velocity.  If we examine the
deviations from this line as function of binary phase, there is a small
amplitude variation (K$\sim$10 km s$^{-1}$), with maximum close to phase
0, as was found by Cowley et al.\ (2002) for He II peaks during the bright
state.  Since our examination of the FES image indicates X0513$-$69 was in
a high optical state during the FUSE observations, it appears that the
narrow O VI 1038\AA\ component probably behaves like the He II 4686\AA\
peaks. 

The 1032\AA\ line is more strongly blended by the interstellar absorption
on its shortward side, and the measures of this line give a scatter about
apparent velocity +380 km s$^{-1}$.  Both line peaks show apparent flux
changes, which are also attributable to blending with airglow lines and
the small velocity variations noted above for the 1038\AA\ line. 

The narrow He II emission at 1084\AA\ is weak and has airglow lines
nearby.  However, we do measure a low velocity amplitude (similar to the
He II 4686\AA\ optical line of Crampton et al.) and a phasing that is
different from the broad O VI, but close to that of the O VI 1038\AA\ peak
mentioned above. 

Table 2 includes our fits to the optical emission peak velocities from
Crampton et al.\ and Southwell et al., for the optically bright and the
transition states, using the new ephemeris.  We note that the FUSE He II
peak velocities, while poorly determined, are quite consistent with the
high-state optical velocity phasing and amplitude.  This is also included
in our binary orbit discussion below.  The He II peak flux shows a small
change which is maximum at phase 0.62 (see Table 2), if the changes are
phase-related. 

Lyman absorptions are clearly present at all phases, and they are seen
well down the series.  The stronger Lyman lines have severe airglow
contamination, but the airglow has a strong decrement, and the lines
measured are free of any significant airglow contamination, as determined
by extrapolation of the decrement. There is a weak correlation of radial
velocity with fraction of night-time in the exposure (hence potential
strength of airglow), but less than half the amplitude with orbital phase,
and there is no measurable change in the equivalent width with phase or
with night-time fraction.  The absorption mean velocity is +226 km
s$^{-1}$, suggesting an outflow from X0513$-$69 whose systemic velocity is
$\sim+280$ km s$^{-1}$ (and not consistent with contamination by zero 
velocity airglow).  We note that the Lyman absorption velocity
changes are closely phased with the increase in broad O VI flux.  If a
broad Lyman emission within the system increases during the same phases, 
then it could conceivably produce a line asymmetry, causing the Lyman absorption
to be measured more positively. However, the Lyman absorptions are only 
1\AA ~wide, compared with 18\AA ~for the broad emission.
We discuss below the puzzle of where the
neutral absorber may be that gives rise to these absorptions, but
draw attention (see Table 2) to the fact that the velocity changes are
close to antiphased with the broad O VI emission velocity. 

\section{A Possible Binary Model} 
    
Here we examine the implications of assuming that the broad O VI emission
velocity curve reflects the orbital motion of the compact star in
X0513$-$69.  Adopting the amplitude and phasing given in Table 2, we have
adopted a weighted mean of the profile-fitting and cross-correlation
measurements, using both the individual spectra and also the phase-binned
spectra of equal signal.  The sketch in Figure 5 shows how the measured
quantities vary with binary phase.  Following Cowley et al.\ (2002), we
assume phase 0 occurs at the superior conjunction of the compact star, and
the phasing of the FUV measures are based on this new ephemeris.  The
broad O VI emission velocity phases are consistent with them showing the
orbital motion of the WD, within their uncertainties (Table 2).  The high
state light curve of Cowley et al.\ has a wide minimum, centered at phase
0 as derived from a sine curve fit to the individual data points.  In the
fainter, ``transition" state the minimum may be narrower, but also
centered on phase 0.  We stress that there is an uncertainty in what is
causing the optical minimum, so that assuming it occurs at conjunction may
be incorrect.  There might instead be an off-centered brighter region that
is partially hidden at phase 0, but it is unlikely to be a structure in
the disk, since it was stable for over 8 years. 

The lack of significant profile changes suggests that the major source of
broad O VI is the inner parts of the accretion disk, rather than in a
stream or the inner part of the Roche lobe where narrow components might
be formed.  By contrast, we note that the O VI emission seen in AM Her, a
highly magnetic CV without a disk, has a complex and varying profile, and
the lines do appear to arise in other parts of the  system where shock
heating may ionize the O VI (Hutchings et al. 2002). 

For X0513$-$69, the adopted semiamplitude of 117 km s$^{-1}$ for the
compact star gives a mass function that implies the values shown in Figure
6.  The lack of any eclipse in the light curve at any wavelength indicates
that the orbital inclination is less than $\sim60^{\circ}$.  The
considerable width of the broad emission argues against a very low orbital
inclination angle.  On the other hand, the observed high velocity of the
bi-polar jets, which presumably are oriented nearly perpendicular to the
disk, is expected to be close to the escape velocity for the white dwarf 
(e.g. Southwell et al. 1996).  At a higher orbital inclination, their 
observed velocities would be expected to be much lower (as seen in the 
supersoft binary RX~J0019.8+2156, Cowley et al.\ 1998). 

These considerations are shown in Figure 6.  The dashed lines show the
locus where the projected jet velocity equals the escape velocity for the
white dwarf, and also the locus of the projected orbital velocity at its
equator.  The jet velocity is assumed not to exceed escape velocity, and
the line width not to exceed orbital velocity at the white dwarf surface.
The diagram shows a region where both conditions apply.  We consider white 
dwarf masses between 0.7 and 1.1M$_{\odot}$, and inclinations above
$\sim$30$^{\circ}$ to be feasible.  We discuss below a mass ratio of
$\sim$2 and a donor star mass of $\sim$1.6 M$_{\odot}$, with an orbital
inclination of 35$^{\circ}$.  Figure 5 shows a sketch of the Roche lobe 
and center of mass for this mass ratio. 

The WD orbital velocity, with an inclination angle $\sim35^{\circ}$, sets
the donor star Roche radius (polar) at $\sim$1.7 R$_{\odot}$.  The main
sequence radius of a star of mass 1.6 M$_{\odot}$ is $\sim$1.2
R$_{\odot}$.  This is also sketched in Figure 5, and it lies well within
the Roche lobe.  On the other hand, a giant (luminosity class III) of this
mass will have a radius of $\sim$4.5 R$_{\odot}$, which would engulf the
companion.  Thus, we suggest that the donor star is just at the end of its
main sequence lifetime and expanding.  This would imply a large mass
transfer rate, and bright disk, as observed.  X0513$-$69 is several times
brighter visually than other supersoft X-ray sources (SSS) in the LMC
(e.g. CAL 83 and CAL 87), although its bolometric luminosity is similar to
that of CAL 83 (L$_{bol}\sim5\times10^{38}$ ergs s$^{-1}$).  Both CAL 83
and X0513$-$69 also show bi-polar jets which are characteristic of a
luminous disk.  The absolute magnitude of a 1.5 M$_{\odot}$ main sequence
star is about +3 (+1.3 for a giant), while the whole X0513$-$69 system has
absolute magnitude of $-2.0$.  Thus, the lack of visibility of the
proposed donor star is not surprising, as it is only a few percent of the
disk luminosity.  Indeed, if the optical light curve is due to heating of
the donor star, as in the case of HZ Her, the observed amplitude is about
that expected for any non-eclipsing inclination. 

 In this model, disk thickening by gas-stream impact is seen from the far
(illuminated) side of the disk at the phase shown for maximum FUV light.  
The He II and O VI peak
velocity probably arises in or near the intersection of the mass stream
and disk, as seen in other X-ray binary systems (e.g. Cyg X-1).  This
material lies near the system center of mass and has low velocity vector
which is the sum of the stream and orbital velocity at this point.  In the
low optical state, when the disk is fainter (smaller?), the relative
change of orbital and streaming velocity vectors leads to the observed
change in phase of maximum peak velocity, as sketched.  The optical
minimum may be caused by the partial occultation of the inner point of the
Roche lobe where the donor star will be brightest.  The He II 1085\AA\
peak flux is maximum at phase 0.62, which is where the heated donor star
is most clearly visible.  However, the feature is weak and near variable
airglow lines. 

The Lyman sharp absorption velocities are antiphased with the broad O VI,
and thus they suggest association with the donor star.  However, the disk
and not the donor star must be the source of the FUV continuum radiation
behind the absorbing material.  Also, as mentioned earlier, the absorption
lines have a mean velocity of +226 km s$^{-1}$ compared with the systemic
velocity of +280 km s$^{-1}$ obtained from the optical He II emission
lines.  Thus, we may be looking through a $\sim$50 km s$^{-1}$ outward
flow of cool material that arises from the donor star and is projected
against the hot continuum of the disk.  The absorbing column does not vary
measurably with binary phase, but the lines are weak and the continuum
does vary.  However the absorptions arise, they may still carry the
orbital motion of the secondary.  It is hard to imagine another way they
could be antiphased with the broad O VI emission.  It is interesting that
this yields a mass ratio of the stars is 54/117=0.46, which lies right in
the region we have already noted in Figure 6.  However, the Lyman
absorption velocity curve is far from sinusoidal, so perhaps the origin of
these lines lies elsewhere. 

The masses suggested here are unusual for SSS in having a more massive
donor star.  As noted, it is possible that the real mass function is
lower, if some of the measured broad velocity changes are due to profile
changes.  However, we have suggested that such a contribution would be
small, so we regard the mass discussion as an upper limit of interest. 
The disk in this system is extremely luminous, and it is widely assumed
that the disk brightness and jet formation occur because the mass transfer
rate is high.  Thus, this is consistent with a more massive donor star. 
As pointed out by Cowley et al.\ (2002), the binary model implied by the
much lower velocity amplitude of the optical line peaks (Crampton et al.,
Southwell et al.) is no longer consistent with the revised ephemeris from
the optical light curve.  Other problems with this original model were the
very low inclination (nearly pole-on) and the low donor-star mass, if one
assumed the compact star was a white dwarf.  

It was originally suggested by van den Heuvel et al. (1992) that the donor
stars in supersoft binary systems would be expected to be about twice the
mass of the white dwarf, based on evolutionary considerations.  However,
until the present data on the FUV O VI velocities, all previous
observational evidence has indicated the donors are low mass stars (see,
for example, review by G\"ansicke et al. 2000). 
As this system appears to differ from other SSS in having a more massive 
donor star, it may be that the mass transfer can be driven by more than
one mechanism. In this case, normal Roche lobe overflow by evolution off 
the main sequence would apply, while for the less massive donor stars,
an additional wind mechanism is required.

\acknowledgments

We are grateful to Alex Fullerton for processing the observations
independently of the regular data pipeline.  APC acknowledges her support 
from NASA and the National Science Foundation.
 
\clearpage
\begin{deluxetable}{cccccccc}
\tablenum{1}
\footnotesize
\tablecaption{Measurements of FUSE Data for RX~J0513.9$-$6951}
\tablehead{
\colhead{Spec.\#} &
\colhead{HJD} &
\colhead{Phase\tablenotemark{a}} &
\colhead{Exp.} &
\multicolumn{2}{c}{Broad O VI Em} &
\colhead{Ly Abs} &
\colhead{Contin } \\
& 
\colhead{2452210+} & &
\colhead{Time} &
\colhead{RV\tablenotemark{b}} &
\colhead{Flux\tablenotemark{c}} &
\colhead{RV\tablenotemark{d}} &
\colhead{1040\AA\tablenotemark{e} }
\\
& 
\colhead{(mid-exp)} &  & 
\colhead{(sec)} &
\colhead{(km s$^{-1}$)} & &
\colhead{(km s$^{-1}$)} & 
}

\startdata

1 & 3.431 & 0.87 & 1571 & 1201  & 7.6 & 200 & 3.6  \nl
2 & 3.502 & 0.97 & 1239 & 1227  & 8.4 & 280 & 4.1 \nl
3 & 3.570 & 0.06 & 973  & 916   & 8.3 & 283 & 4.2 \nl
4 & 3.611 & 0.11 & 367  & 930   & 10.5 & 290 & 6.4 \nl
5 & 3.643 & 0.16 & 918  & 963   & 9.3 & 350 & 4.3 \nl
6 & 3.684 & 0.21 & 801  & 965   & 7.0 & 313 & 5.2 \nl
7 & 3.714 & 0.25 & 650  & 740   & 8.8 & 291 & 4.9 \nl
8 & 3.755 & 0.30 & 1230 & 934   & 6.5 & 240 & 4.4 \nl
9 & 3.827 & 0.40 & 1610 & 1147  & 6.2 & 214 & 4.0 \cr
10 & 3.906 & 0.50 & 3240 & 1153 & 4.9 & 185 & 3.1 \nl
11 & 3.976 & 0.59 & 3310 & 1217 & 4.8 & 197 & 3.0 \nl
12 & 4.045 & 0.68 & 3298 & 1156 & 5.3 & 196 & 3.0 \nl
13 & 4.114 & 0.77 & 3281 & 1026 & 5.1 & 186 & 3.1 \nl
14 & 4.184 & 0.86 & 3281 & 1190 & 4.9 & 184 & 3.2 \nl
15 & 4.257 & 0.96 & 2571 & 1099 & 5.6 & 229 & 3.3 \nl
16 & 4.328 & 0.05 & 1832 & 1159 & 6.4 & 166 & 3.4 \nl

\enddata
\tablenotetext{a}{Ephemeris: T$_0$ = JD 2448858.099 + 0.7629434E days, where
T$_0$ is the time of $V_{MACHO}$ minimum light (see Cowley et al. 2002)}  
\tablenotetext{b}{$\pm$40 km s$^{-1}$}
\tablenotetext{c}{Flux units: $\times10^{-13}$ ergs s$^{-1}$ cm$^{-2}$}
\tablenotetext{d}{$\pm$20 km s$^{-1}$}
\tablenotetext{e}{Mean Value: units $\times10^{-14}$ ergs s$^{-1}$ cm$^{-2}$}
\end{deluxetable}

\begin{deluxetable}{cccccc}
\tablenum{2}
\footnotesize
\tablecaption{Phasing of FUV Spectroscopic Variations in RX~J0513.9$-$6951}
\tablehead{
\colhead{Data} &
\colhead{HJD} &
\colhead{Semiamplitude} &
\colhead{Mean Error} &
\colhead{Binary} &
\colhead{Sine Fit} \\ 
  & \colhead{2452210+}& 
\colhead{K} &  &
\colhead{Phase \tablenotemark{a}} &
\colhead{Mean} \\
 \colhead{\bf{Radial Velocity}} & 
\colhead{(Time of Max.) } &
\colhead{(km s$^{-1}$)} &
\colhead{(km s$^{-1}$)} &  & 
\colhead{(km s$^{-1}$)}
}

\startdata
O VI Mean em \tablenotemark{b} & ~ 4.00 $\pm$0.05 & ~ 126 $\pm$55 & ~ 152 &
~ 0.62 $\pm$0.07 & ~ 1141 $\pm$41 \cr
O VI CCor em & ~ 4.07 $\pm$0.04 & ~ 110 $\pm$36 & ~ 94 & ~ 0.71 $\pm$0.06 &
~ 1094 $\pm$26 \cr
Binned spec, mean  &~4.06$\pm$0.03 &~121$\pm$30 &~64 &~0.70$\pm$0.05 
&~1118$\pm$22\cr
Adopt && ~ 117 && ~ 0.70 \cr
H Lyman abs & ~ 3.64 $\pm$0.02 & ~ 54 $\pm$10 & ~ 23 & ~ 0.15 $\pm$0.02 & ~
226 $\pm$7 \cr
He II em & ~ 3.56 $\pm$0.19 & ~ 11 $\pm$ 11 & ~ 22 & ~ 0.05 $\pm$0.25 & ~ 268
$\pm$7 \cr
\hline
\bf{Line Flux\tablenotemark{c}}
&& ~ ($\times10^{-13}$) & ~ ($\times10^{-13}$) &  & ~ ($\times10^{-13}$) \nl
\hline
O VI & ~ 3.63 $\pm$0.02 & ~ 2.0 $\pm$0.4 & ~ 1.2 & ~ 0.14 $\pm$0.03 &
~ 6.4 $\pm$0.3 \cr
He II & ~ 4.00 $\pm$0.06 & ~ 0.27 $\pm$0.15 & ~ 0.30 & ~ 0.62 $\pm$0.08
& ~ 0.46 $\pm$0.01 \cr
\hline
\bf{O VI Eq.W} && ~ (\AA) & ~ (\AA) && ~ (\AA) \nl
\hline
Mean \tablenotemark{b} & ~ 3.51 $\pm$0.05 & ~ 2.3 $\pm$0.9 & ~ 3.0 & ~
0.15 $\pm$0.06 & ~ 17 $\pm$1 \cr
\hline
\bf{Continuum\tablenotemark{c}}  
& ~ (2452210+) & ~ ($\times10^{-14}$) & ~ ($\times10^{-14}$) &  & ~
($\times10^{-14}$) \nl
\hline
920-965\AA\ & ~ 3.71 $\pm$0.04 & ~ 2.0 $\pm$0.4 & ~ 1.1 & ~ 0.24
$\pm$0.03 & ~ 7.0 $\pm$0.3 \cr
1050-1100\AA\ & ~ 3.69 $\pm$0.03 & ~ 1.5 $\pm$0.3 & ~ 0.9 & ~ 0.22
$\pm$0.03 & ~ 5.6 $\pm$0.2 \cr
1100-1180\AA\ & ~ 3.68 $\pm$0.02 & ~ 1.3 $\pm$0.3 & ~ 0.7 & ~ 0.20
$\pm$0.03 & ~ 4.2 $\pm$0.2 \cr
\hline
\bf{Opt He II peaks} &2449660+\cr
\hline
high opt state  &9.93$\pm$0.02 &11.2$\pm$2.2 &7.5 &0.08$\pm$0.03 &279$\pm$2\cr
intermediate opt state
 &9.78$\pm$0.03 &11.8$\pm$2.4 &8.9 &0.88$\pm$0.04 &299$\pm$2\cr

\enddata

\tablenotetext{a}{Ephemeris: T$_0$ = JD 2448858.099 + 0.7629434E days, where
T$_0$ is the time of $V_{MAC}$ minimum light (see Cowley et al. 2002)}  
\tablenotetext{b}{Mean of profile fits to LiF1A, LiF2B, SiC2A}
\tablenotetext{c}{Flux Units: ergs s$^{-1}$ \AA$^{-1}$ cm$^{-2}$}

\end{deluxetable}

\clearpage

\begin{figure}
\caption{Mean FUSE spectrum of X0513$-$69.  Each panel is from one FUSE
spectral channel, as identified, so the wavelength scales differ and
overlap somewhat.  Emission features are airglow lines, except for He II
and O VI, as marked.  Some of the Lyman absorption lines that arise in the
binary system are identified, and the positions of the non-detected C III
and N III lines are shown.  All identifications are shifted to the
LMC velocity frame.} 
\end{figure}

\begin{figure}
\caption{Measured quantities in X0513$-$69 plotted against orbital 
phase.  Typical error bars are shown for the velocity measurements.  The 
O VI velocities are the average of cross-correlation and profile-fitting
measures and are shown for all spectra (solid points) and for spectra
binned for equal signal (open points).  The continuum values in the lower
panel are fitted continuum levels and not the total signal as in Table 1. 
Although the variations are not sinusoidal, O VI emissions and Lyman
absorptions appear to vary in antiphase.  The orbital phases are based on
the new Cowley et al.\ (2002) ephemeris.} 
\end{figure}

\begin{figure}
\caption{The uppermost plot show the mean O VI doublet profile with
airglow and peak emission shown as edited out.  The position of the O VI
lines with a LMC velocity are marked.  The airglow emissions are sharper
than shown here in the heavily smoothed plot, so that the contamination by
airglow is less severe than the plot suggests.  The absorption lines are
interstellar.  Below the mean spectrum are individual spectrum differences
from the mean, edited of airglow, plotted by phase.  These are derived from
the set of 9 phase-binned spectra.  Overall changes are seen to be smooth
and symmetrical across the profile center, indicating that there are no
broad profile shape changes.} 
\end{figure}

\begin{figure}
\caption{Mean O VI profile folded in velocity space about the mean LMC
wavelength of the doublet.  There appears to be a strong P Cygni absorption
underneath a broad emission which reaches to some 4000 km s$^{-1}$, 
causing the large apparent wavelength shift of the emission.  There are weak
peaks from each of the O VI lines that are blended with airglow emission
and interstellar absorptions.} 
\end{figure}

\begin{figure}
\caption{Sketch of the binary system as proposed, with the key measured
quantity variations indicated.  The text discusses where the various
spectral features appear to arise.  The circle inside the donor star shows
the main sequence radius of a star of the mass discussed.  Lower panel 
suggests the viewing cone, if the orbital inclination angle is 
$i=35^{\circ}$. } 
\end{figure}

\begin{figure}
\caption{Mass diagram for the binary system for various orbital
inclinations, from the broad O VI velocities.  The likely range of values
is indicated by the dashed lines, which are the limits defined by the jet
velocity and broad line width.  The asterisk is the value discussed
in detail in the text.} 
\end{figure}

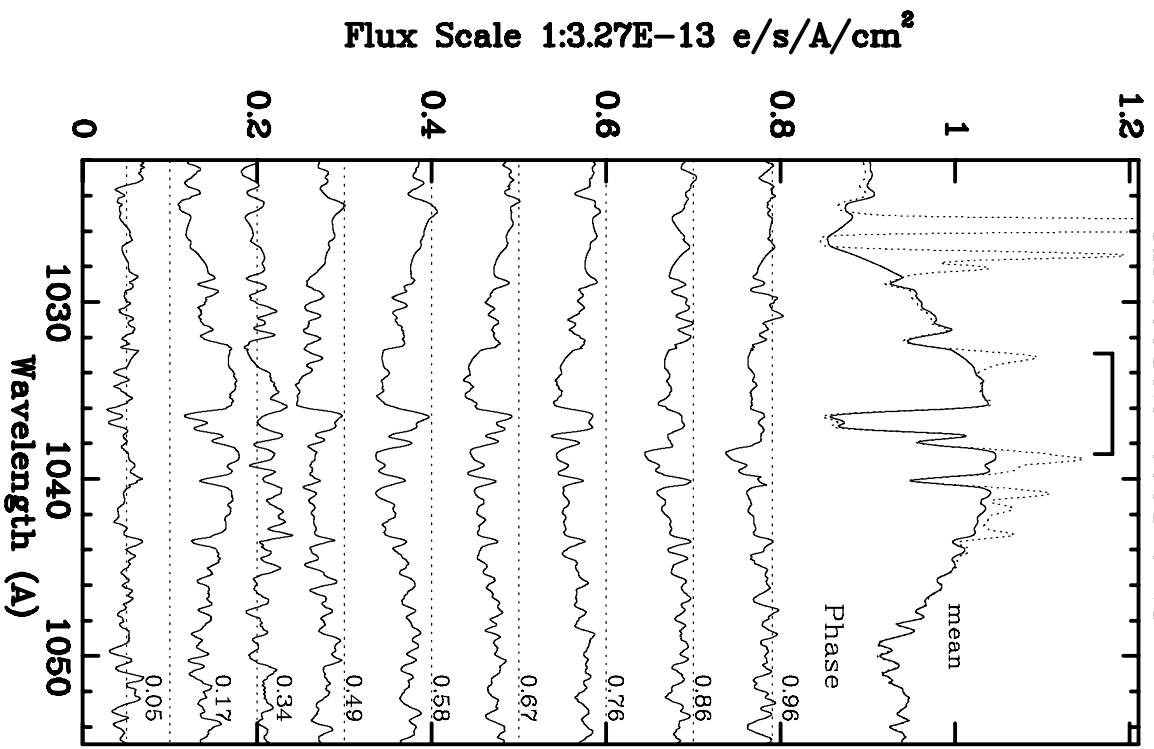
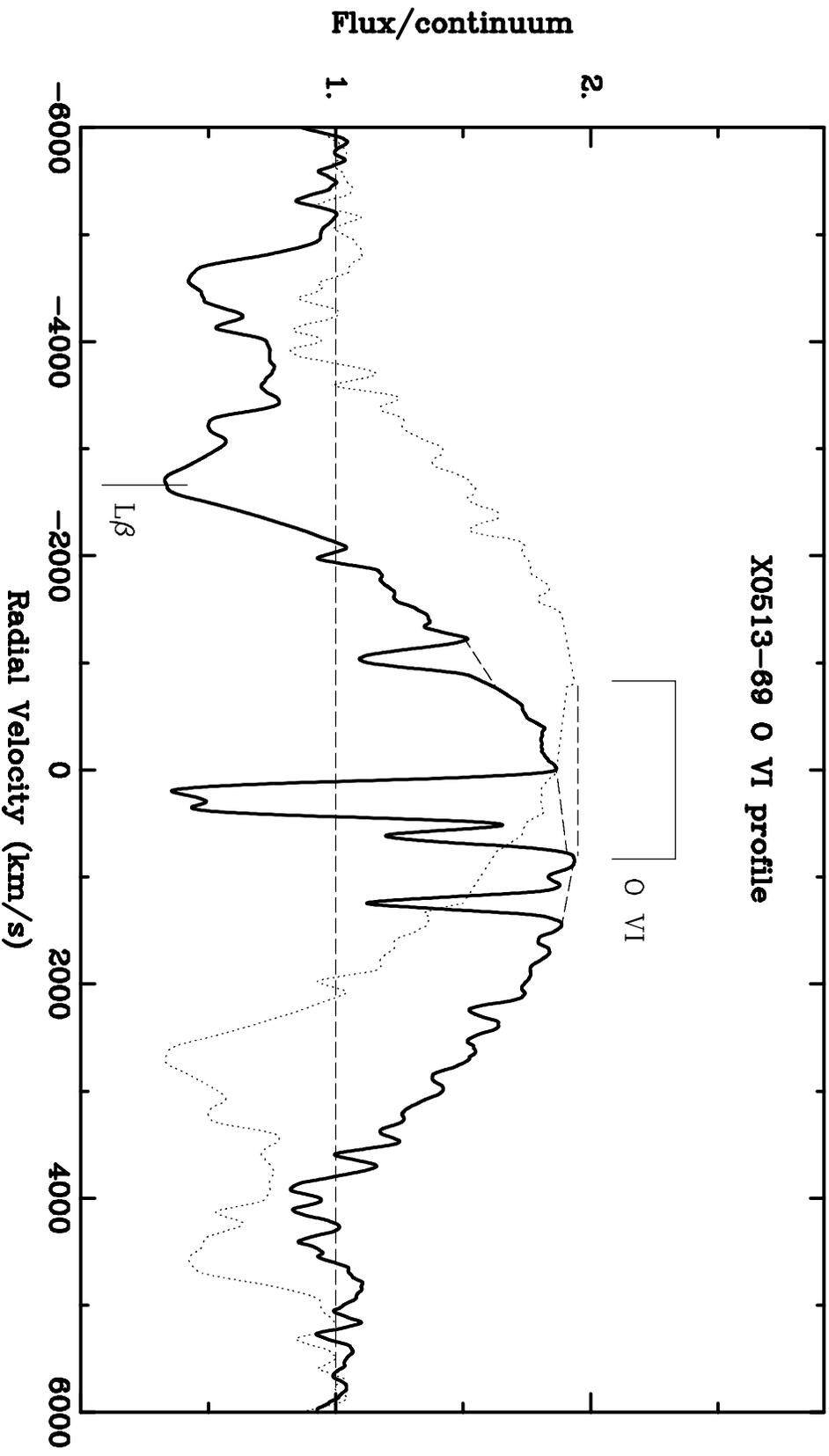
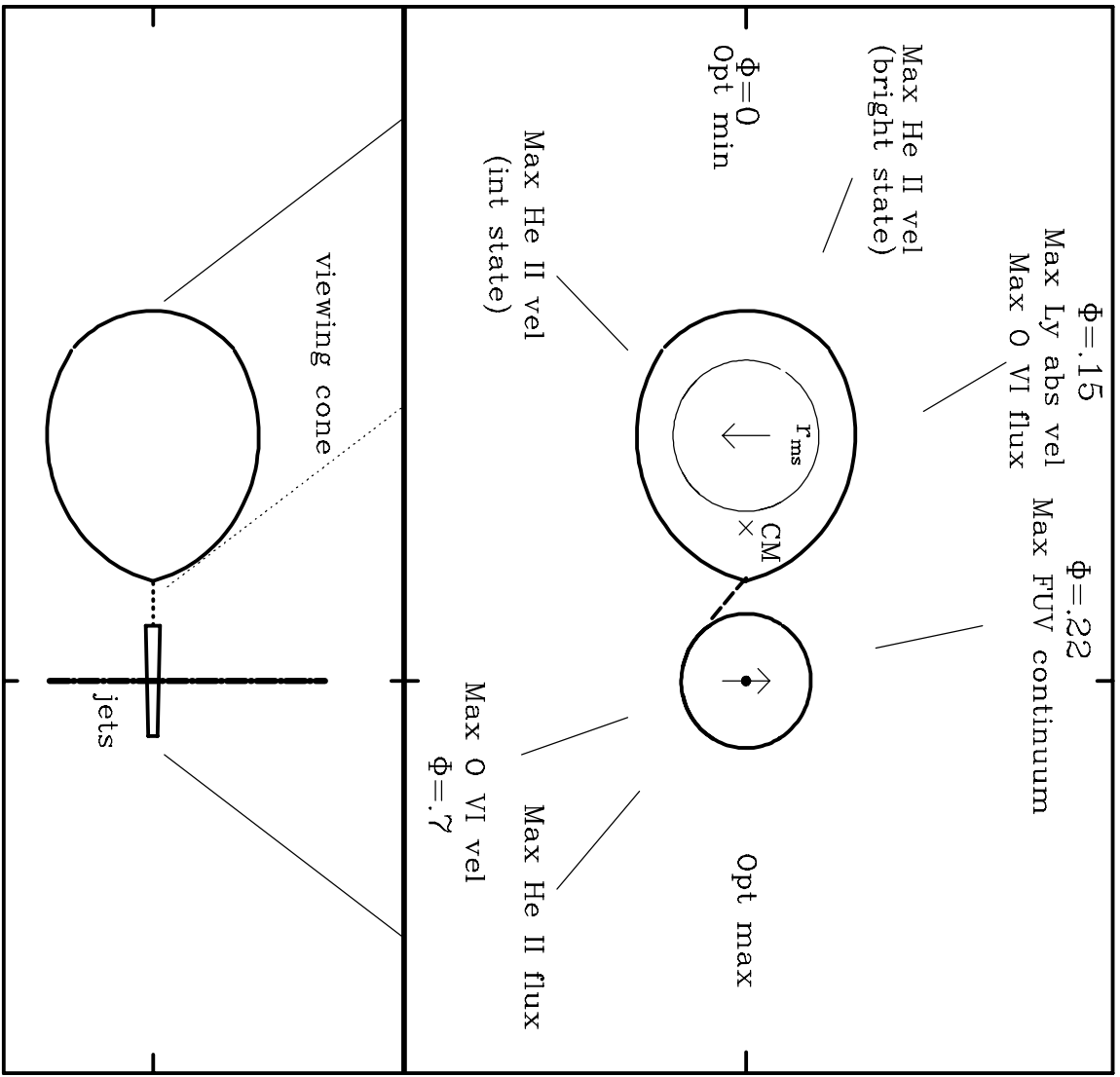
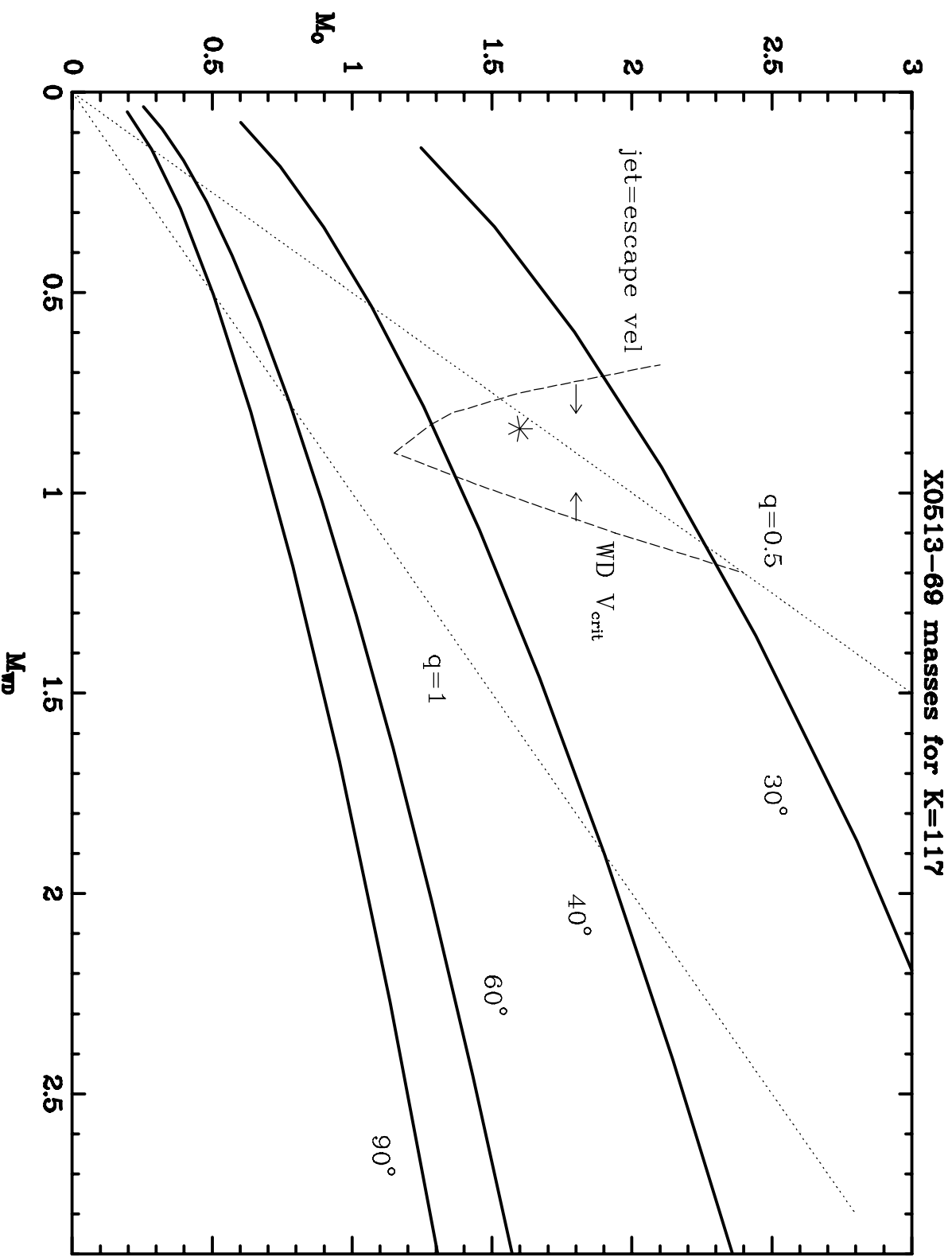
\end{document}